Quantum-chemical studies of rutile nanoparticles toxicity I. Defect-free rod-like model clusters

Martin Breza

Department of Physical Chemistry, Slovak Technical University, Radlinskeho 9, SK-81237 Bratislava, Slovakia

e-mail: martin.breza@stuba.sk

Abstract

Using semiempirical PM6 method, the structures of a rod-like $[Ti_{40}O_{124}H_{81}]^{-7}$ model cluster and of the $[Ti_{40}O_{124}H_{81}Cu]^{-5}$ ones with $Cu^{2+}$ coordinated at various sites were optimized. If the relative toxicity of individual Ti centers in rod-like rutile nanoparticles can be evaluated by the electron density transfer to a $Cu^{2+}$ probe, its maximal values may be ascribed to the pentacoordinated corner and hexacoordinated edge ones with three Ti-OH bonds. However, these centers exhibit the least negative interaction energies which can be compensated by the significantly better accessibility of the corner Ti center in comparison with the remaining ones. The Ti centers with the most negative interaction energy parameters exhibit the lowest extent of the electron density transfer to a $Cu^{2+}$ probe. The rutile nanoparticles destruction starts at pentacoordinated Ti face centers

Keywords: protonated rutile nanoparticle; PM6 semiempirical method; Cu(II) probe; electron density transfer; active centers

Introduction

Nano-sized $TiO_2$ particles can be found in a large number of foods and consumer products. Their nanotoxicity has been drawn an increasing attention because human bodies are potentially exposed to this nanomaterial either by inhalation, oral or dermal route. Numerous studies have tried to characterize their in vivo biodistribution, clearance and toxicological effects, especially in in lungs, liver, kidneys, spleen, brain, lymph nodes, testis, blood and lungs of rats (see e.g. [1-7]).

Rutile is the most stable polymorph of $TiO_2$ at all temperatures exhibiting lower total free energy than metastable phases of anatase or brookite [8]. Rutile has a tetragonal unit cell (space group $P4_2/mnm$) [9]. The titanium cations are surrounded by an octahedron of 6 oxygen atoms. The oxygen anions have a coordination number of 3, resulting in a trigonal planar coordination. Rutile crystals are most commonly observed to exhibit a prismatic or acicular growth habit with preferential orientation along their c axis, the [001] direction. This growth habit is favored as the {110} facets of rutile exhibit the lowest surface free energy and are therefore thermodynamically most stable.[10]

The interaction of water with $TiO_2$ is crucial to many of its practical applications .The rutile (110)–aqueous solution interface structure was measured [11] in deionized water (DIW) and 1 molal (m) RbCl + RbOH solution (pH 12) at 25 ºC with the X-ray crystal truncation rod method. The rutile surface in both solutions consists of a stoichiometric (1 : 1) surface unit mesh with the surface terminated by bridging oxygen (BO) and terminal oxygen (TO) sites, with a mixture of water molecules and hydroxyl groups (OH) occupying the TO sites. An additional hydration layer is observed above the TO site, with three distinct water adsorption sites each having well-defined vertical and lateral locations. Structural displacements of atoms at the oxide surface are sensitive to the solution composition. Ti atom displacements from their bulk lattice positions, as large as 0.05 Å at the rutile (110)–DIW interface, decay in magnitude into the crystal with significant relaxations that are observable down to the fourth Ti-layer below the surface. A systematic outward shift was observed for Ti atom locations below the BO rows, while a systematic inward displacement was found for Ti atoms below the TO rows. The Ti displacements were mostly reduced in contact with the RbCl solution at pH 12, with no statistically significant relaxations in the fourth layer Ti atoms. The distance between the surface 5-fold Ti atoms and the oxygen atoms of the TO site is 2.13 ± 0.03 Å in DIW and 2.05 ± 0.03 Å in the $Rb^+$ solution, suggesting molecular adsorption of water at the TO site to the rutile (110) surface in DIW, while at pH 12, adsorption at the TO site is primarily in the form of an adsorbed hydroxyl group.

Very recently, scanning tunnelling microscopy and surface X-ray diffraction were used [12] to determine the structure of the rutile (110)–aqueous interface, which is comprised of an ordered array of hydroxyl molecules with molecular water in the second layer. A combination of data from real-space imaging, spectroscopic measurements and surface X-ray diffraction, with interpretation aided by DFT calculations implies that the rutile TiO2(110) surface has terminal hydroxyls in the contact layer. The ideal coverage by terminal OH groups is half a monolayer, but this is decreased to approximately 0.4 monolayers by absences at domain wall boundaries.

According to Alagona and Ghio [13, 14] the antioxidant activity of prenylated pterocarpans is to be related to their copper coordination ability. Based on B3LYP calculations of several complexes with $Cu^{2+}$ of their low-energy conformers their metal ion affinity (MIA) values have been determined. In aqueous solution, the solvent effect dampens the free energy differences and reduces the MIA especially when the ion is remarkably exposed to the solvent The stability order of the metalated species at the various coordination sites strongly depends on their position and nature. The spin density of the cation upon ligand coordination becomes vanishingly small, whereas the ligand spin density approaches 1. Thus the ligand is oxidized to a radical cation ($Ligand^{\bullet+}$), while Cu(II) is reduced to Cu(I). In agreement with experimental investigations, the higher antioxidant activity of individual compounds and their reaction sites may be assigned to higher MIA values and higher reducing character toward Cu(II). The antioxidant ability of various sites of hyperjovinol A through their ability to coordinate a $Cu^{2+}$ ion and reduce it to $Cu^+$ was successfully tested by Mammino [15] as well. Another modification of the above mentioned method has been used for both N centers of a series of para-phenylene diamine (PPD) antioxidants. [16] Nearly linear dependence of the experimental antioxidant effectiveness on Cu(II)-PPD interaction energies, Cu atomic charges and other electron density parameters has been deduced.

From the chemical point of view, the nanoparticles toxicity is based on an electron density transfer to human tissues as well. Therefore the above mentioned method, which has been tested on antioxidants, might be suitable for the relative toxicity estimation of various sites of model nanoparticles. The liquids in human body are of aqueous character which implies a protonation of the negative charged surface of the rutile nanoparticles. For the sake of simplicity, only hexacoordinated Ti atoms and full protonation of non-bridging O atoms may be supposed in model systems. Semiempirical methods of quantum chemistry seem to be a suitable compromise between molecular mechanics methods, which are suitable for large model systems but say nothing on their electron structure, and DFT methods, which bring valuable information on electron distribution within the studied systems but their size is significantly restricted due to technical reasons. The aim of this study is to estimate toxicity of various sites of an idealized protonated rod-like rutile nanoparticle (over 200 atoms) based on its Cu(II) complexation ability and electron density transfer to Cu at semiempirical PM6 level of theory.

Method

Geometries of the model systems under study were optimized using semiempirical PM6 method of quantum chemistry [17]. Stability of the optimized structures was confirmed by vibrational analysis (no imaginary vibrations). Atomic charges were evaluated in terms of Mulliken population analysis (MPA) [18] and alternatively atomic polar tensor (APT) derived charges [19]. All the calculations were performed using Gaussian09 program package [20].

The metal-ligand interaction energy $\Delta_{int}E$ is defined as

$$\Delta_{int}E = E_{Complex} - E_{Ligand} - E_{ion} \tag{1}$$

where $E_{Complex}$ and $E_{Ligand}$ are the energies of the $[L...Cu]^{q+2}$ complex and of the isolated rutile nanoparticle $L^q$ model cluster in their optimized geometries, respectively, and $E_{Ion}$ is the energy of the isolated $Cu^{2+}$ ion [13-15]. Analogously metal-ligand interaction enthalpy $\Delta_{int}H_{298}$ and Gibbs free energy $\Delta_{int}G_{298}$ at 298 K data were evaluated as well.

The deformation energy $E_{def}$ is the difference [13-15] between the energy of the ligand $L^q$ in each complex geometry (Ligand(Cu)) and that corresponding to its optimized structure (Ligand(opt))

$$E_{def} = E_{Ligand(Cu)} - E_{Ligand(opt)} \tag{2}$$

Deformation energies should be smaller than the corresponding metal-ligand ones.

Results and discussion

Using experimental rutile structure [21], in the first step we have formed an idealized rod-like $[Ti_{40}O_{124}]^{-88}$ cluster (Fig. 1) of ca 1.5 nm x 1 nm x 1nm size. Its planes are parallel with the (1 1 0) plane of the rutile unit cell and all Ti atoms are hexacoordinated. As such a highly negative nanoparticle cannot exist in biological water-based environment, all monovalent O atoms have been protonated to form a $[Ti_{40}O_{124}H_{81}]^{-7}$ cluster. Its geometry has been optimized in the singlet ground spin state (Fig. 2). We may see that its planes are significantly deformed due to protonation and the original Ti hexacoordination is sometimes reduced to pentacoordination. We can distinguish several Ti centres according to their bonding to hydroxyl groups (OH) and bridging oxygens ($O_b$) between two Ti atoms. These centers may be divided into 3 groups as follows:

A. At the rod corners only pentacoordinated $Ti(OH)_3(O_b)_2$ centers may be found

B. The rod edges contain either hexacoordinated $Ti(OH)_3(O_b)_3$ (B1), $Ti(OH)_2(O_b)_4$ (B2) and $Ti(OH)(O_b)5$ (B3) centers or pentacoordinated $Ti(OH)_4(O_b)$ (B4), $Ti(OH)_3(O_b)_2$ (see model B5) and $Ti(OH)_2(O_b)_3$ (B6) centers.

C. The rod faces have hexacoordinated $Ti(OH)(O_b)_5$ (C1) or pentacoordinated $Ti(OH)(O_b)_4$ (C2) centers.

In order to compare the reactivity of all the above mentioned possible reaction sites, we have added a $Cu^{2+}$ ion at the distance of ca 1.9 – 2.3 Å from the hydroxyl groups of every center under study. The geometries of in this way created $[Ti_{40}O_{124}H_{81}Cu]^{-5}$ clusters were again PM6 optimized in the ground doublet spin state. The resulting structures may be seen in Figs. 3 - 6. For the sake of simplicity, the notation of the above reaction sites agrees with the $[Ti_{40}O_{124}H_{81}Cu]^{-5}$ model system labels. Their selected characteristics are presented in Tables 1 and 2 and in Table S1 of Supplementary information.

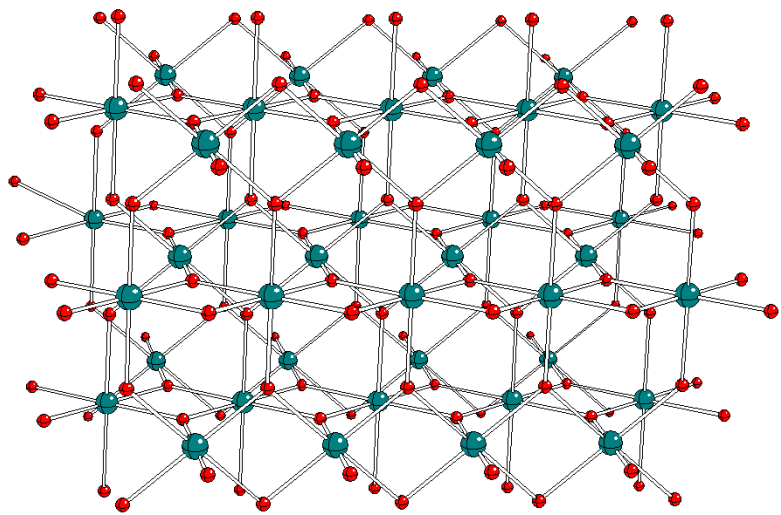

Fig. 1. $[Ti_{40}O_{124}]^{-88}$ cluster in experimental rutile geometry [21] (Ti – green, O – red).

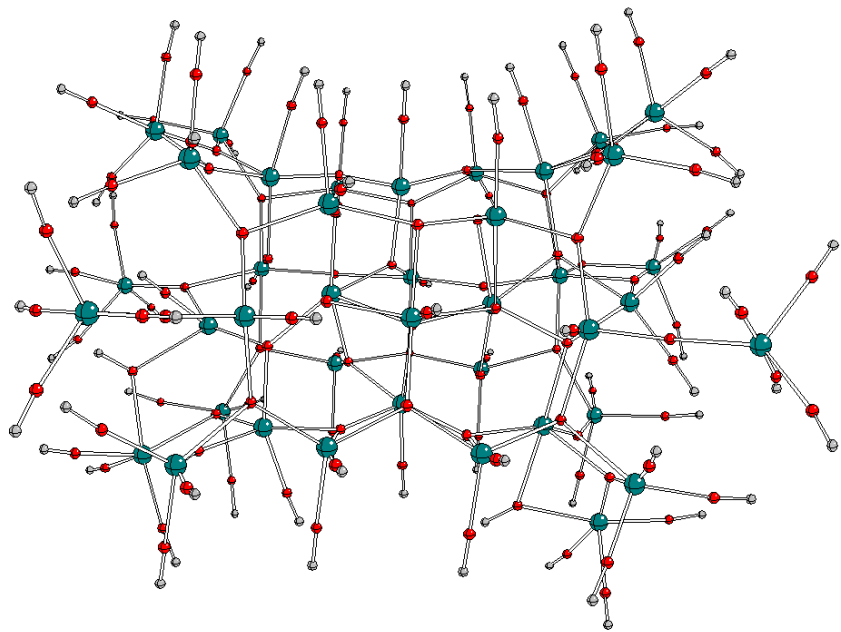

Fig. 2 PM6 optimized geometry of a $[Ti_{40}O_{124}H_{81}]^{-7}$ cluster (Ti – green, O – red, H - grey).

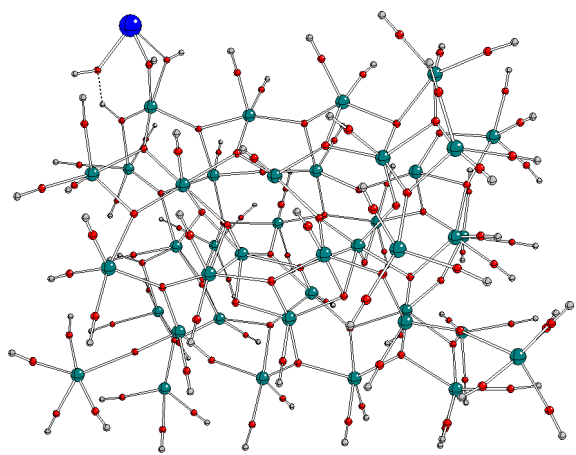

Fig. 3 PM6 optimized geometry of a [Ti$_{40}$O$_{124}$H$_{81}$Cu]$^{-5}$ cluster, model A (Ti – green, O – red, H – grey, Cu blue).

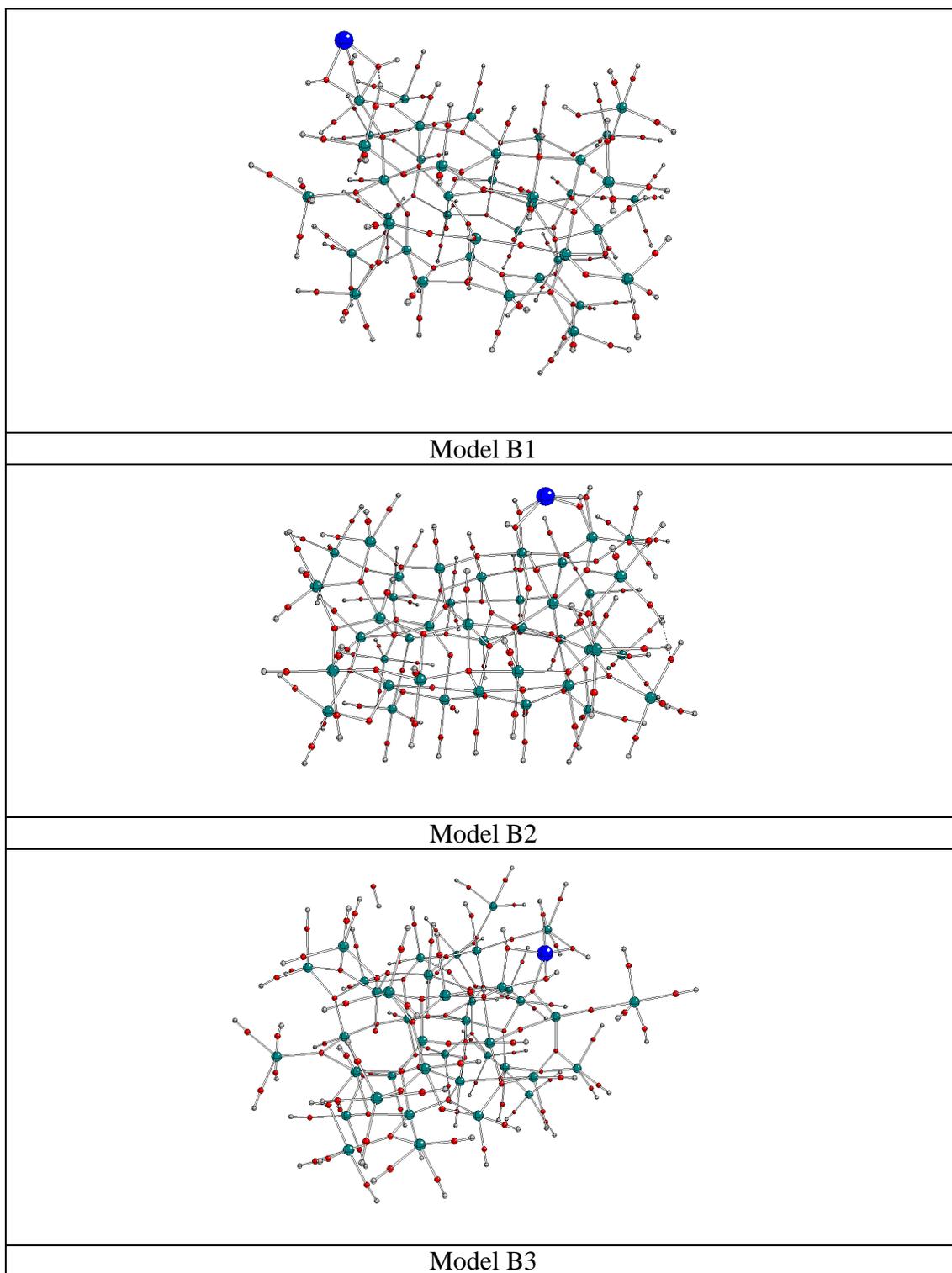

Fig. 4 PM6 optimized geometry of $[Ti_{40}O_{124}H_{81}Cu]^{-5}$ clusters, models B1 – B3 (see Fig. 3 for atom notation)

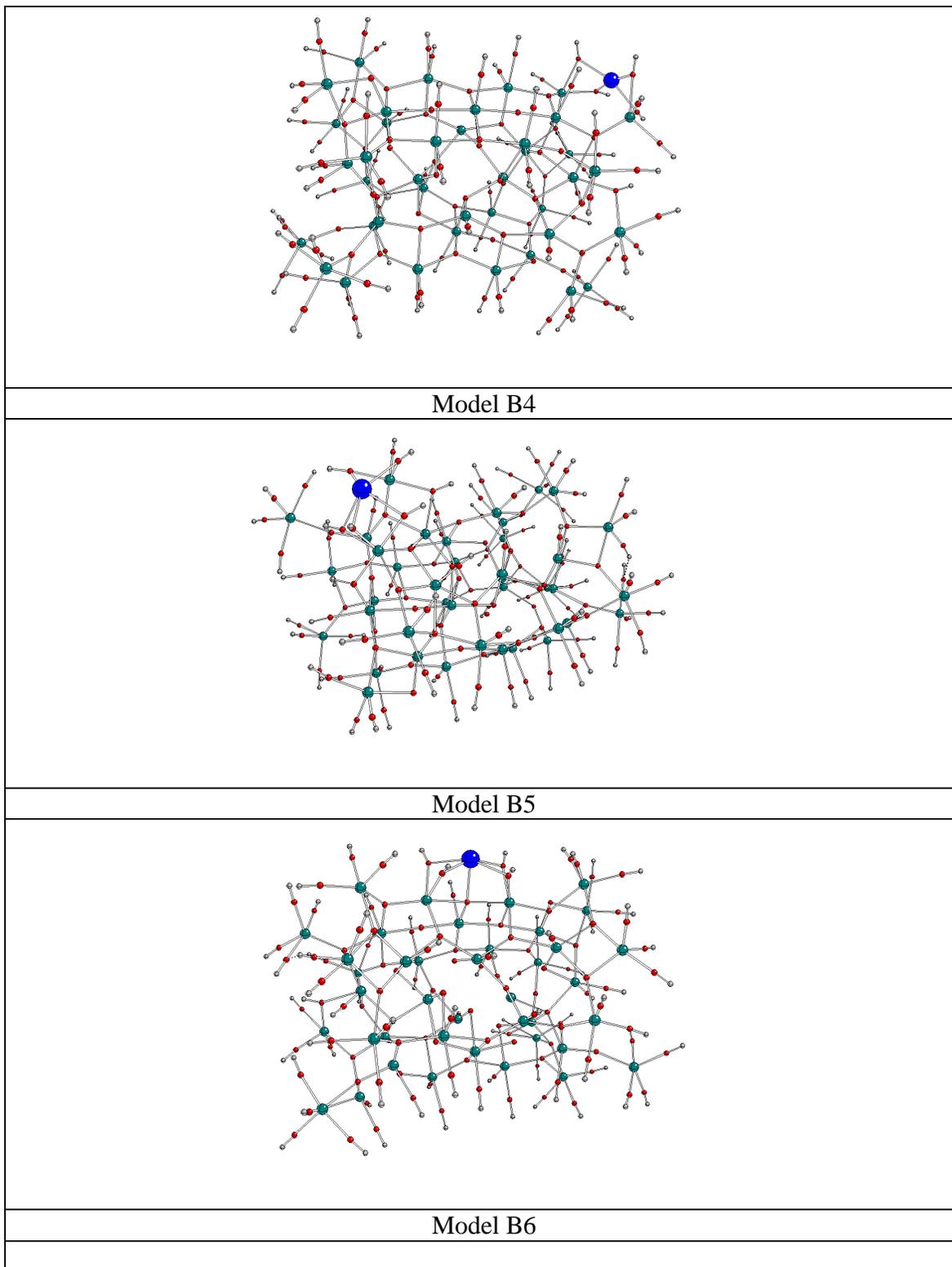

Fig. 5 PM6 optimized geometry of [Ti$_{40}$O$_{124}$H$_{81}$Cu]$^{-5}$ clusters, models B4 – B6 (see Fig. 3 for atom notation).

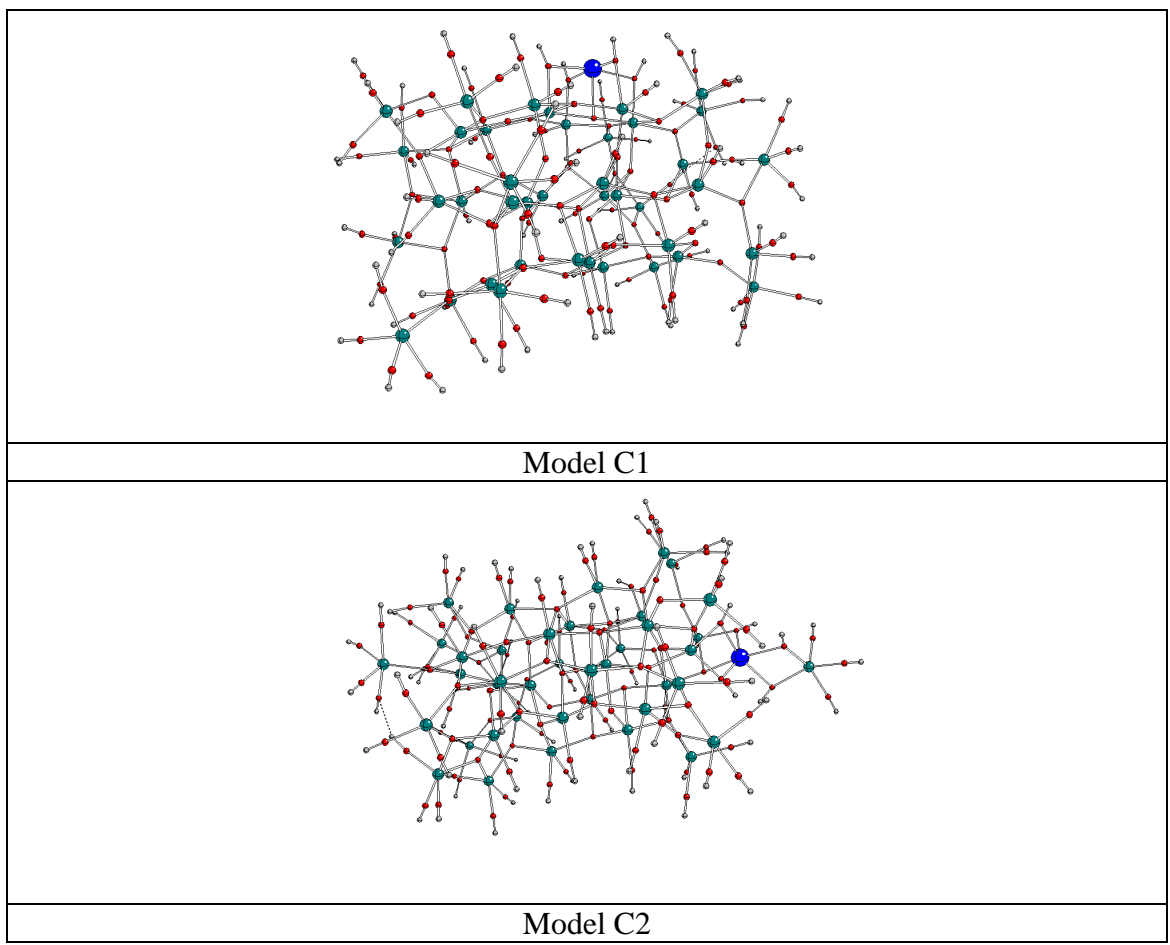

Fig. 6 PM6 optimized geometry of [$Ti_{40}O_{124}H_{81}Cu$]$^{-5}$ clusters, models C1 and C2 (see Fig. 3 for atom notation).

Table 1. Copper(II)-ligand interaction energies ($\Delta_{int}E$), Gibbs free interaction energies ($\Delta_{int}G_{298}$) and interaction enthalpies ($\Delta_{int}H_{298}$) at 298 K, and deformation energies ($E_{def}$) of $^2$[Ti$_{40}$O$_{124}$H$_{81}$Cu]$^{5-}$ structures obtained by PM6 method for the model systems under study.

| Model | $\Delta_{int}E$ [kcal/mol] | $\Delta_{int}G_{298}$ [kcal/mol] | $\Delta_{int}H_{298}$ [kcal/mol] | $E_{def}$ [kcal/mol] |
|---|---|---|---|---|
| B A | -905.8 | -886.1 | -901.2 | 289.9 |
| C B1 | -929.4 | -909.1 | -923.2 | 285.9 |
| D B2 | -967.2 | -945.3 | -960.6 | 313.7 |
| E B3 | -1001.5 | -982.5 | -996.0 | 334.1 |
| A B4 | -976.7 | -954.5 | -970.0 | 323.6 |
| F B5 | -967.3 | -944.1 | -961.5 | 318.2 |
| G B6 | -951.5 | -929.3 | -945.7 | 332.8 |
| H C1 | -968.7 | -947.6 | -963.0 | 343.7 |
| I C2 | -992.5 | -969.8 | -986.3 | 332.8 |

Our results (Table 1) indicate that all copper(II)-ligand interaction energy parameters exhibit the same trends. Their differences between various model systems are higher than the corresponding differences in deformation energies. Thus no corrections in the observed trends of the calculated interaction energy are necessary. As expected, the corner Ti centers (model A) exhibit the least negative interaction energy parameters. The most negative values have the hexacoordinated edge Ti centers with single hydroxyl group (model B3). These are shifted to less negative values with increasing number of hydroxyl groups in other hexacoodinated edge Ti centers (B2 and B1 models). Pentacoordinated edge Ti centers (B4, B5 and B6 models) exhibit reverse trends. Interaction energy parameters of hexacoordinated face Ti centers (C1 model) are comparable with the medians of the edge Ti centers. Their values for pentacoodinated face Ti centers (C2 model) are significantly more negative due to the removal of [Ti(OH)$_5$]$^-$ cluster from the nanoparticle by a Cu$^{2+}$ probe. This indicates the most probable site of a nanoparticle destruction.

Table 2. MPA (q(Cu)$_{MPA}$) and APT (q(Cu)$_{APT}$) copper atomic charges and the lengths of Cu-O bond to hydroxyl (d$_{Cu-OH}$) and bridging (d$_{Cu-Obr}$) oxygen atoms of $^2$[Ti$_{40}$O$_{124}$H$_{81}$Cu]$^{5-}$ structures obtained by PM6 method for the model systems under study.

| Model | q(Cu)$_{MPA}$ | q(Cu)$_{APT}$ | d$_{Cu-OH}$ [Å]$^{a)}$ | d$_{Cu-Obr}$ [Å] |
|---|---|---|---|---|
| B A | 0.586 | 0.669 | 2.009, 2.038, 2.048 (2.032) | - |
| C B1 | 0.585 | 0.638 | 2.018, 2.034, 2.039 (2.030) | - |
| D B2 | 0.633 | 0.721 | 2.068, 2.077, 2.086 (2.077) | - |
| E B3 | 0.648 | 0.744 | 2.089, 2.124, 2.126, 2.184 (2.131) | - |
| A B4 | 0.634 | 0.734 | 2.040, 2.058, 2.062, 2.105 (2.066) | - |
| F B5 | 0.648 | 0.721 | 2.126, 2.130, 2.138, 2.160, 2.193 (2.149) | - |
| G B6 | 0.645 | 0.740 | 2.133, 2.139, 2.158, 2.198 (2.157) | 2.107 |
| H C1 | 0.627 | 0.739 | 2.081, 2.120, 2.137, 2.152 (2.123) | 2.146 |
| I C2 | 0.641 | 0.899 | 2.141, 2.177, 2.194, 2.256, 2.294 (2.212) | 2.071 |

Remarks:

$^{a)}$Averaged values in parentheses

Table 2 contains charges of Cu probes indicating the extent of an electron density transfer from the ligand (the higher Cu charge corresponds to the lower electron density transfer). Resulting Cu spin density is vanishing in all the model systems under study and thus the corresponding data are not presented. Positive APT charges at Cu atoms are higher than the MPA ones and need not exhibit the same trends for pentacoordinated centers. In general. more negative interaction energy parameters (Table 1) are connected with higher Cu charges but these are significantly affected by the number and type of Cu bonded oxygen atoms (OH or O$_b$). Cu probes in A, B1 and B2 models are tricoordinated, in B3 and B4 models tetracoordinated and in the B5 one pentacoordinated by hydroxyls. The remaining models have the Cu probe coordinated by a single bridging oxygen atom and 4 – 5 hydroxyls which might cause irregularities in Cu charges. Averaged Cu-OH bond lengths (as a measure of corresponding Cu-O bond strengths implied by an electron density transfer) follow the trends in interaction energy parameters and Cu charges for hexacoordinated Ti edge centers unlike the pentacoordinated ones.

If the relative toxicity of individual Ti centers in rod-like rutile nanoparticles can be evaluated by the electron density transfer to a $Cu^{2+}$ probe, its maximal values may be ascribed to the pentacoordinated corner (A model) and hexacoordinated edge ones with three Ti-OH bonds (B1 model). However, these centers exhibit the least negative interaction energies which can be compensated by the significantly better accessibility of the corner Ti center (A model) in comparison with the remaining ones. The Ti centers with the most negative interaction energy parameters exhibit the lowest extent of the electron density transfer to a $Cu^{2+}$ probe (B3 an C2 models). The rutile nanoparticles destruction starts at pentacoordinated Ti face centers (C2 model).

Our model systems consist of three $TiO_2$ planes only and their protonation causes too large planes warping in comparison with significantly larger real systems. As quantum-chemical calculations of larger model systems are connected with serious technical problems even at semiempirical level of theory, an ONIOM (our own n-layered integrated molecular orbital and molecular mechanics) treatment [22] combining semiempirical and molecular mechanics calculations must be used. Molecular mechanics methods enable to increase the size of model systems (which in our case reduce the plane warping due to surface protonation) but say nothing on their electron structure. In our future studies the above mentioned conclusions on toxicity of individual Ti centers in rutile nanoparticles will be tested on model systems of various shapes and sizes.

Acknowledgements

Financial support of H2020-NMP-2014-2015/H2020-NMP-2015-two-stage project No. 685817 (HISENTS) is appreciated. We thank the HPC Center at the Slovak University of Technology in Bratislava, which is a part of the Slovak Infrastructure of High Performance Computing (SIVVP Project, ITMS Code 26230120002, funded by the European Region Development Funds), for computing facilities.